# PREDICTING STUDENTS' PERFORMANCE USING ID3 AND C4.5 CLASSIFICATION ALGORITHMS

Kalpesh Adhatrao, Aditya Gaykar, Amiraj Dhawan, Rohit Jha and Vipul Honrao

Department of Computer Engineering,
Fr. C.R.I.T., Navi Mumbai, Maharashtra, India

*ABSTRACT*

*An educational institution needs to have an approximate prior knowledge of enrolled students to predict their performance in future academics. This helps them to identify promising students and also provides them an opportunity to pay attention to and improve those who would probably get lower grades. As a solution, we have developed a system which can predict the performance of students from their previous performances using concepts of data mining techniques under Classification. We have analyzed the data set containing information about students, such as gender, marks scored in the board examinations of classes X and XII, marks and rank in entrance examinations and results in first year of the previous batch of students. By applying the ID3 (Iterative Dichotomiser 3) and C4.5 classification algorithms on this data, we have predicted the general and individual performance of freshly admitted students in future examinations.*

## 1. INTRODUCTION

Every year, educational institutes admit students under various courses from different locations, educational background and with varying merit scores in entrance examinations. Moreover, schools and junior colleges may be affiliated to different boards, each board having different subjects in their curricula and also different level of depths in their subjects. Analyzing the past performance of admitted students would provide a better perspective of the probable academic performance of students in the future. This can very well be achieved using the concepts of data mining.

For this purpose, we have analysed the data of students enrolled in first year of engineering. This data was obtained from the information provided by the admitted students to the institute. It includes their full name, gender, application ID, scores in board examinations of classes X and XII, scores in entrance examinations, category and admission type. We then applied the ID3 and C4.5 algorithms after pruning the dataset to predict the results of these students in their first semester as precisely as possible.

## 2. LITERATURE SURVEY

### 2.1. Data Mining

Data mining is the process of discovering interesting knowledge, such as associations, patterns, changes, significant structures and anomalies, from large amounts of data stored in databases or data warehouses or other information repositories [1]. It has been widely used in recent years due to the availability of huge amounts of data in electronic form, and there is a need for turning such data into useful information and knowledge for large applications. These applications are found in fields such as Artificial Intelligence, Machine Learning, Market Analysis, Statistics and Database Systems, Business Management and Decision Support [2].

#### 2.1.1. Classification

Classification is a data mining technique that maps data into predefined groups or classes. It is a supervised learning method which requires labelled training data to generate rules for classifying test data into predetermined groups or classes [2]. It is a two-phase process. The first phase is the learning phase, where the training data is analyzed and classification rules are generated. The next phase is the classification, where test data is classified into classes according to the generated rules. Since classification algorithms require that classes be defined based on data attribute values, we had created an attribute "class" for every student, which can have a value of either "Pass" or "Fail".

#### 2.1.2. Clustering

Clustering is the process of grouping a set of elements in such a way that the elements in the same group or cluster are more similar to each other than to those in other groups or clusters [1]. It is a common technique for statistical data analysis used in the fields of pattern recognition, information retrieval, bioinformatics, machine learning and image analysis. Clustering can be achieved by various algorithms that differ about the similarities required between elements of a cluster and how to find the elements of the clusters efficiently. Most algorithms used for clustering try to create clusters with small distances among the cluster elements, intervals, dense areas of the data space or particular statistical distributions.

### 2.2. Selecting Classification over Clustering

In clustering, classes are unknown apriori and are discovered from the data. Since our goal is to predict students' performance into either of the predefined classes - "Pass" and "Fail", clustering is not a suitable choice and so we have used classification algorithms instead of clustering algorithms.

### 2.3. Issues Regarding Classification

#### 2.3.1. Missing Data

Missing data values cause problems during both the training phase and to the classification process itself. For example, the reason for non-availability of data may be due to [2]:

- Equipment malfunction
- Deletion due to inconsistency with other recorded data

- Non-entry of data due to misunderstanding
- Certain data considered unimportant at the time of entry
- No registration of data or its change

This missing data can be handled using following approaches [3]:

- Data miners can ignore the missing data
- Data miners can replace all missing values with a single global constant
- Data miners can replace a missing value with its feature mean for the given class
- Data miners and domain experts, together, can manually examine samples with missing values and enter a reasonable, probable or expected value

In our case, the chances of getting missing values in the training data are very less. The training data is to be retrieved from the admission records of a particular institute and the attributes considered for the input of classification process are mandatory for each student. The tuple which is found to have missing value for any attribute will be ignored from training set as the missing values cannot be predicted or set to some default value. Considering low chances of the occurrence of missing data, ignoring missing data will not affect the accuracy adversely.

**2.3.2. Measuring Accuracy**

Determining which data mining technique is best depends on the interpretation of the problem by users. Usually, the performance of algorithms is examined by evaluating the accuracy of the result. Classification accuracy is calculated by determining the percentage of tuples placed in the correct class. At the same time there may be a cost associated with an incorrect assignment to the wrong class which can be ignored.

**2.4. ID3 Algorithm**

In decision tree learning, ID3 (Iterative Dichotomiser 3) is an algorithm invented by Ross Quinlan used to generate a decision tree from the dataset. ID3 is typically used in the machine learning and natural language processing domains. The decision tree technique involves constructing a tree to model the classification process. Once a tree is built, it is applied to each tuple in the database and results in classification for that tuple. The following issues are faced by most decision tree algorithms [2]:

- Choosing splitting attributes
- Ordering of splitting attributes
- Number of splits to take
- Balance of tree structure and pruning
- Stopping criteria

The ID3 algorithm is a classification algorithm based on Information Entropy, its basic idea is that all examples are mapped to different categories according to different values of the condition attribute set; its core is to determine the best classification attribute form condition attribute sets. The algorithm chooses information gain as attribute selection criteria; usually the attribute that has the highest information gain is selected as the splitting attribute of current node, in order to make information entropy that the divided subsets need smallest [4]. According to the different values of the attribute, branches can be established, and the process above is recursively called on

41

each branch to create other nodes and branches until all the samples in a branch belong to the same category. To select the splitting attributes, the concepts of Entropy and Information Gain are used.

### 2.4.1. Entropy

Given probabilities $p_1, p_2, \ldots, p_s$, where $\sum p_i = 1$, Entropy is defined as

$$H(p_1, p_2, \ldots, p_s) = \sum - (p_i \log p_i)$$

Entropy finds the amount of order in a given database state. A value of H = 0 identifies a perfectly classified set. In other words, the higher the entropy, the higher the potential to improve the classification process.

### 2.4.2. Information Gain

ID3 chooses the splitting attribute with the highest gain in information, where gain is defined as difference between how much information is needed after the split. This is calculated by determining the differences between the entropies of the original dataset and the weighted sum of the entropies from each of the subdivided datasets. The formula used for this purpose is:

$$G(D, S) = H(D) - \sum P(D_i) H(D_i)$$

## 2.5. C4.5

C4.5 is a well-known algorithm used to generate a decision trees. It is an extension of the ID3 algorithm used to overcome its disadvantages. The decision trees generated by the C4.5 algorithm can be used for classification, and for this reason, C4.5 is also referred to as a statistical classifier. The C4.5 algorithm made a number of changes to improve ID3 algorithm [2]. Some of these are:

- Handling training data with missing values of attributes
- Handling differing cost attributes
- Pruning the decision tree after its creation
- Handling attributes with discrete and continuous values

Let the training data be a set $S = s_1, s_2 \ldots$ of already classified samples. Each sample $S_i = x_1, x_2\ldots$ is a vector where $x_1, x_2 \ldots$ represent attributes or features of the sample. The training data is a vector $C = c_1, c_2\ldots$, where $c_1, c_2\ldots$ represent the class to which each sample belongs to.

At each node of the tree, C4.5 chooses one attribute of the data that most effectively splits data set of samples S into subsets that can be one class or the other [5]. It is the normalized information gain (difference in entropy) that results from choosing an attribute for splitting the data. The attribute factor with the highest normalized information gain is considered to make the decision. The C4.5 algorithm then continues on the smaller sub-lists having next highest normalized information gain.

## 3. TECHNOLOGIES USED

### 3.1. HTML and CSS

HyperText Markup Language (HTML) is a markup language for creating web pages or other information to display in a web browser. HTML allows images and objects to be included and that can be used to create interactive forms. From this, structured documents are created by using structural semantics for text such as headings, links, lists, paragraphs, quotes etc.

CSS (Cascading Style Sheets) is designed to enable the separation between document content (in HTML or similar markup languages) and document presentation. This technique is used to improve content accessibility also to provide more flexibility and control in the specification of content and presentation characteristics. This enables multiple pages to share formatting and reduce redundancies.

### 3.2. PHP and the CodeIgniter Framework

PHP (recursive acronym for PHP: Hypertext Preprocessor) is a widely-used open source general-purpose server side scripting language that is especially suited for web development and can be embedded into HTML.

CodeIgniter is a well-known open source web application framework used for building dynamic web applications in PHP [6]. Its goal is to enable developers to develop projects quickly by providing a rich set of libraries and functionalities for commonly used tasks with a simple interface and logical structure for accessing these libraries. CodeIgniter is loosely based on the Model-View-Controller (MVC) pattern and we have used it to build the front end of our implementation.

### 3.3. MySQL

MySQL is the most popular open source RDBMS which is supported, distributed and developed by Oracle [8]. In the implementation of our web application, we have used it to store user information and students' data.

### 3.4. RapidMiner

RapidMiner is an open source data mining tool that provides data mining and machine learning procedures including data loading and transformation, data preprocessing and visualization, modelling, evaluation, and deployment [7]. It is written in the Java programming language and makes use of learning schemes and attribute evaluators from the WEKA machine learning environment and statistical modelling schemes for the R-Project. We have used RapidMiner to generate decision trees of ID3 and C4.5 algorithms.

## 4. IMPLEMENTATION

We had divided the entire implementation into five stages. In the first stage, information about students who have been admitted to the second year was collected. This included the details submitted to the college at the time of enrolment. In the second stage, extraneous information was removed from the collected data and the relevant information was fed into a database. The third stage involved applying the ID3 and C4.5 algorithms on the training data to obtain decision trees of both the algorithms. In the next stage, the test data, i.e. information about students currently enrolled in the first year, was applied to the decision trees. The final stage consisted of developing the front end in the form of a web application.

These stages of implementation are depicted in Figure 1.

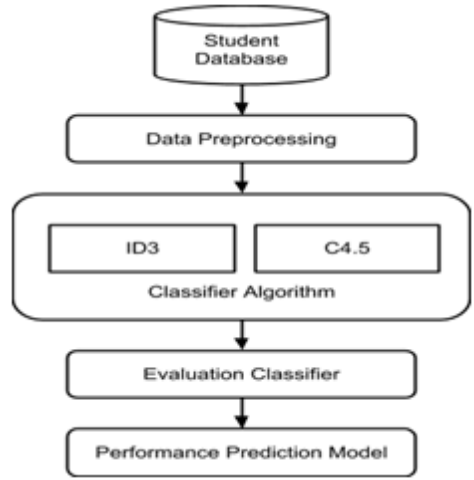

Figure 1. Processing model

## 4.1. Student Database

We were provided with a training dataset consisting of information about students admitted to the first year. This data was in the form of a Microsoft Excel 2003 spreadsheet and had details of each student such as full name, application ID, gender, caste, percentage of marks obtained in board examinations of classes X and XII, percentage of marks obtained in Physics, Chemistry and Mathematics in class XII, marks obtained in the entrance examination, admission type, etc. For ease of performing data mining operations, the data was filled into a MySQL database.

## 4.2. Data Preprocessing

Once we had details of all the students, we then segmented the training dataset further, considering various feasible splitting attributes, i.e. the attributes which would have a higher impact on the performance of a student. For instance, we had considered 'location' as a splitting attribute, and then segmented the data according to students' locality.

A snapshot of the student database is shown in Figure 2. Here, irrelevant attributes such as students residential address, name, application ID, etc. had been removed. For example, the admission date of the student was irrelevant in predicting the future performance of the student. The attributes that had been retained are those for merit score or marks scored in entrance examination, gender, percentage of marks scored in Physics, Chemistry and Mathematics in the board examination of class XII and admission type. Finally, the "class" attribute was added and it held the predicted result, which can be either "Pass" or "Fail".

Since the attributes for marks would have discrete values, to produce better results, specific classes were defined. Thus, the "merit" attribute had a value "good" if the merit score of the student was 120 or above out of a maximum score of 200, and was classified as "bad" if the merit score was below 120. Also, the value that can be held by the "percentage" attribute of the student are three - "distinction" if the percentage of marks scored by the student in the subjects of Physics, Chemistry and Mathematics was 70 or above, "first_class" if the percentage was less than 70 and greater than or equal to 60, then it was classified as "second_class" if the percentage was less than 60. The attribute for admission type is labelled "type" and the value held by a student for it can be either "AI" (short for All-India), if the student was admitted to a seat available for All-India candidates, or "OTHER" if the student was admitted to another seat.

| sr_no | merit_no | merit_marks | app_id | name | gender | cast | location | percent | type |
|---|---|---|---|---|---|---|---|---|---|
| 1 | 328 | 153.00 | EN10205034 | AKSHAY DEBNATH | Male | Open | Mumbai | 95.66 | AI |
| 2 | 725 | 152.00 | EN10279070 | YEMPALLE SUSHMA BASWARAJ | Female | Open | Mumbai | 86.66 | AI |
| 3 | 1066 | 143.00 | EN10288911 | KIRAN SUSHIL GRIFFITHS | Male | Open | Mumbai | 96.00 | AI |
| 4 | 1294 | 136.00 | EN10167854 | WALCHALE ABHIJEET SUHAS | Male | Open | Mumbai | 82.00 | AI |
| 5 | 1419 | 132.00 | EN10255786 | KUNAL JADHAV | Male | Open | Mumbai | 80.33 | AI |
| 6 | 21566 | 109.00 | EN10230782 | KARKHELE RAVINDRAKUMAR VITTHAL | Male | NT 3 (NT-D) | Mumbai | 83.66 | GNT3H |
| 7 | 3290 | 156.00 | EN10172564 | TALAWADEKAR ADITYA SHYAM | Male | OBC | Mumbai | 89.33 | GOBCH |
| 8 | 5933 | 144.00 | EN10264877 | SONAWANE NIKHIL RAJENDRA | Male | SBC/OBC | Mumbai | 89.66 | GOBCH |
| 9 | 6882 | 140.00 | EN10196064 | PATIL SUMEET BHAGWAN | Male | OBC | Mumbai | 88.33 | GOBCH |
| 11 | 1456 | 168.00 | EN10195904 | LOHOTE PRANIT TANAJI | Male | Open | Mumbai | 92.00 | GOPENH |
| 12 | 2158 | 162.00 | EN10216545 | IYER SIDDHARTH SUNDARAM | Male | Open | Mumbai | 93.66 | GOPENH |
| 13 | 2519 | 160.00 | EN10255191 | GEORGE NISHANT JOSEPH | Male | Open | Mumbai | 94.66 | GOPENH |

| merit | gender | percent | type | class |
|---|---|---|---|---|
| good | Male | distinction | AI | pass |
| good | Female | distinction | AI | pass |
| good | Male | distinction | AI | pass |
| good | Male | distinction | AI | pass |
| good | Male | distinction | AI | pass |
| bad | Male | distinction | OTHER | pass |
| good | Male | distinction | OTHER | pass |
| good | Male | distinction | OTHER | pass |
| good | Male | distinction | OTHER | fail |
| good | Male | distinction | OTHER | pass |
| good | Male | distinction | OTHER | pass |

Figure 2. Preprocessed student database

## 4.3. Data Processing Using RapidMiner

The next step was to feed the pruned student database as input to RapidMiner. This helped us in evaluating interesting results by applying classification algorithms on the student training dataset. The results obtained are shown in the following subsections:

### 4.3.1. ID3 Algorithm

Since ID3 is a decision tree algorithm, we obtained a decision tree as the final result with all the splitting attributes and it is shown in Figure 3.

International Journal of Data Mining & Knowledge Management Process (IJDKP) Vol.3, No.5, September 2013

### 4.3.2. C4.5 Algorithm

The C4.5 algorithm too generates a decision tree, and we obtained one from RapidMiner in the same way as ID3. This tree, shown in Figure 4, has fewer decision nodes as compared to the tree for improved ID3, which is shown in Figure 3.

## 4.4. Implementing the Performance Prediction Web Application

RapidMiner helped significantly in finding hidden information from the training dataset. These newly learnt predictive patterns for predicting students' performance were then implemented in a working web application for staff members to use to get the predicted results of admitted students.

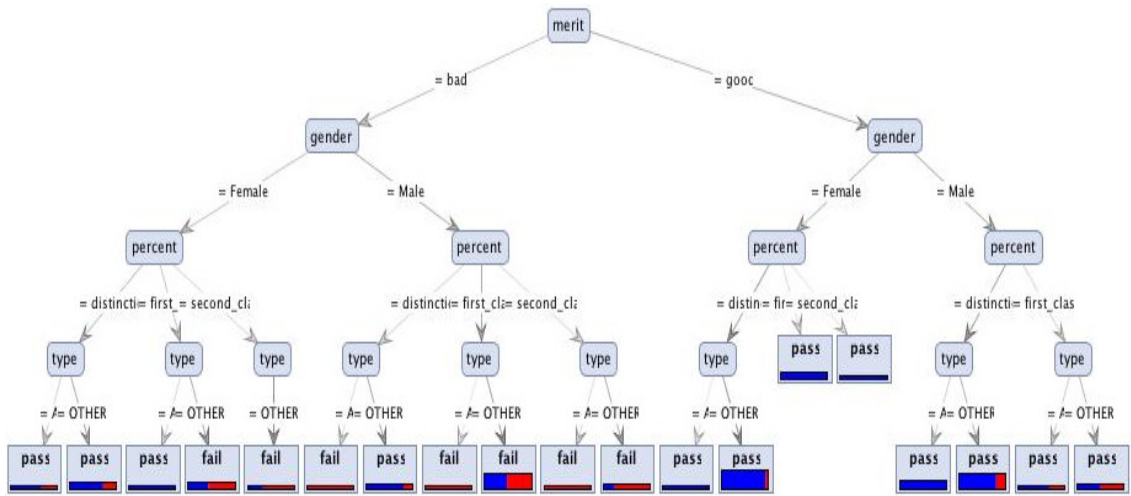

Figure 3. Decision tree for ID3

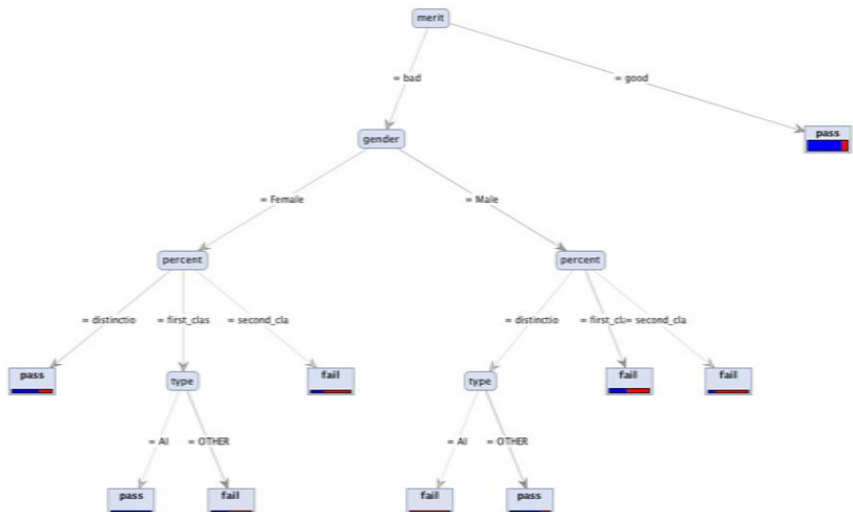

Figure 4. Decision tree for C4.5

### 4.4.1. CodeIgniter

The web application was developed using a popular PHP framework named CodeIgniter. The application has provisions for multiple simultaneous staff registrations and staff logins. This ensures that the work of no two staff members is interrupted during performance evaluation. Figure 5 and Figure 6 depict the staff registration and staff login pages respectively.

### 4.4.2. Mapping Decision Trees to PHP

The essence of the web application was to map the results achieved after data processing to code. This was done in form of class methods in PHP. The result of the improved ID3 and C4.5 algorithms were in the form of trees and these were translated to code in the form of if-else ladders. We then placed these ladders into PHP class methods that accept only the splitting attributes - PCM percentage, merit marks, admission type and gender as method parameters. The class methods return the final result of that particular evaluation, indicating whether that student would pass or fail in the first semester examination. Figure 7 shows a class method with the if-else ladder.

Figure 5. Registration page for staff members

Figure 6. Login page for staff members

International Journal of Data Mining & Knowledge Management Process (IJDKP) Vol.3, No.5, September 2013

**4.4.3. Singular Evaluation**

Once the decision trees were mapped as class methods, we built a web page for staff members to feed values for the name, application ID and splitting attributes of a student, as can be seen in Figure 8. These values were then used to predict the result of that student as either "Pass" or "Fail".

**4.4.4. Upload Excel Sheet**

Singular Evaluation is beneficial when the results of a small number of students are to be predicted, one at a time. But in case of large testing datasets, it is feasible to upload a data file in a format such as that of a Microsoft Excel spreadsheet, and evaluate each student's record. For this, staff members can upload a spreadsheet containing records of students with attributes in a predetermined order. Figure 9 shows the upload page for Excel spreadsheets.

```php
public function dtalgo3($percent, $merit, $ad_type, $gender){
    if( $percent == "distinction" )
        return "pass";
    else{
        if( $percent == "first_class" ){
            if( $merit == "bad" ){
                if( $ad_type == "AI" )
                    return "pass";
                else
                    return "fail";
            }
            else
                return "pass";
        }
        else
            return "fail";
    }
}
```

Figure 7.  PHP class method mapping a decision tree

Figure 8.  Web page for Singular Evaluation

International Journal of Data Mining & Knowledge Management Process (IJDKP) Vol.3, No.5, September 2013### 4.4.5. Bulk Evaluation

Under the Bulk Evaluation tab, a staff member can choose an uploaded dataset to evaluate the results, along with the algorithm to be applied over it. After submitting the dataset and algorithm, the predicted result of each student is displayed in a table as the value of the attribute "class". A sample result of Bulk Evaluation can be seen in Figure 10.

Figure 9. Page to upload Excel spreadsheet

| merit_marks | app_id | name | gender | caste | location | percent | type | class |
|---|---|---|---|---|---|---|---|---|
| 153 | DX10205034 | AKSHAY DEBNATH | Male | Open | Mumbai | 95.66 | AI | PASS |
| 152 | DX10279070 | YEMPALLE SUSHMA BASWARAJ | Female | Open | Mumbai | 86.66 | AI | PASS |
| 143 | DX10288911 | KIRAN SUSHIL GRIFFITHS | Male | Open | Mumbai | 96 | AI | PASS |
| 136 | DX10167854 | WALCHALE ABHIJEET SUHAS | Male | Open | Mumbai | 82 | AI | PASS |
| 132 | DX10255786 | KUNAL JADHAV | Male | Open | Mumbai | 80.33 | AI | PASS |
| 109 | DX10230782 | KARKHELE RAVINDRAKUMAR VITTHAL | Male | NT 3 (NT-D) | Mumbai | 83.66 | GNT3H | PASS |
| 156 | DX10172564 | TALAWADEKAR ADITYA SHYAM | Male | OBC | Mumbai | 89.33 | GOBCH | PASS |

Figure 10. Page showing results after Bulk Evaluation

### 4.4.6. Verifying Accuracy of Predicted Results

The accuracy of the algorithm results can be tested under the Verify tab. A staff member has to select the uploaded verification file which already has the actual results and the algorithm that has to be tested for accuracy. After submission the predicted result of evaluation is compared with actual results obtained and the accuracy is calculated. Figure 11 shows that the accuracy achieved is 75.145% for both ID3 and C4.5 algorithms. Figure 12 shows the mismatched tuples, i.e. the tuples which were predicted wrongly by the application for the current test data.

49

International Journal of Data Mining & Knowledge Management Process (IJDKP) Vol.3, No.5, September 2013

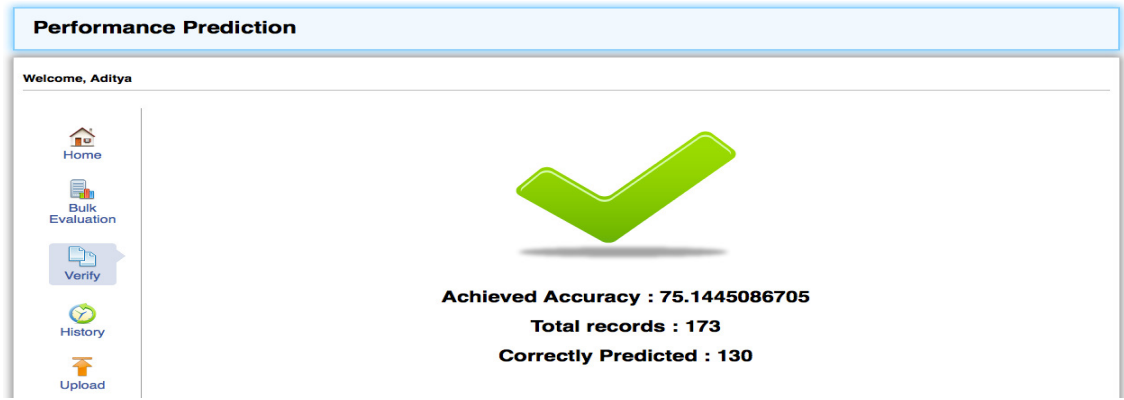

Figure 11. Accuracy achieved after evaluation

| merit_marks | app_id | name | gender | caste | location | percent | type | class | Predicted |
|---|---|---|---|---|---|---|---|---|---|
| 140 | DX10196064 | PATIL SUMEET BHAGWAN | Male | OBC | Mumbai | 88.33 | GOBCH | fail | PASS |
| 124 | DX10297565 | MAHAJAN NISHANT VIJAY | Male | OBC | North Maharashtra | 58 | GOBCO | fail | PASS |
| 118 | DX10356072 | NARKHEDE JUHI RAJEEV | Female | Open | North Maharashtra | 76.33 | LOPENO | pass | FAIL |
| 108 | DX10149595 | WAGHAMARE LAXMAN PANDURANG | Male | OBC | Shivaji + Solapur | 74 | GOBCO | pass | FAIL |
| 153 | DX10182982 | JAISWAL ABHAY SHAILESH | Male | Open | Mumbai | 75.66 | GOPENH | fail | PASS |
| 150 | DX10193225 | RAJPUT ABHISHEK DANSINGH | Male | Open | Mumbai | 82 | GOPENH | fail | PASS |
| 93 | DX10260441 | RAMYA MACHERI | Female | Open | Mumbai | 73 | AI | pass | FAIL |

Figure 12. Mismatched tuples shown during verification

### 4.4.7. Singular Evaluation History

Using the web interface, staff members can view all Singular Evaluations they had conducted in the past. This is displayed in the form of a table, containing attributes of the student and the predicted result. If required, a record from this table may be deleted by a staff member. A snapshot of this table is shown in Figure 13.

| Application ID | Name | Gender | Percentage | Merit marks | Admission Type | Algorithm | Class | |
|---|---|---|---|---|---|---|---|---|
| DX123456 | Aditya Gaykar | Male | 89.17 | 157 | OTHER | C4.5 | pass | Delete |
| DX123456 | Rahul | Male | 123 | 89 | OTHER | Decision Tree | pass | Delete |
| DX121312 | Aditya Gaykar | Male | 90.33 | 157 | OTHER | Decision Tree | pass | Delete |

Figure 13. History of Singular Evaluations performed by staff members

## 5. FUTURE WORK

In this project, prediction parameters such as the decision trees generated using RapidMiner are not updated dynamically within the source code. In the future, we plan to make the entire implementation dynamic to train the prediction parameters itself when new training sets are fed into the web application. Also, in the current implementation, we have not considered extra-curricular activities and other vocational courses completed by students, which we believe may have a significant impact on the overall performance of the students. Considering such parameters would result in better accuracy of prediction.

## 6. CONCLUSIONS

In this paper, we have explained the system we have used to predict the results of students currently in the first year of engineering, based on the results obtained by students currently in the second year of engineering during their first year.

The results of Bulk Evaluation are shown in Table 1. Random test cases considered during individual testing resulted in approximately equal accuracy, as indicated in Table 2.

Table 1. Results of Bulk Evaluation

| Algorithm | Total Students | Students whose results are correctly predicted | Accuracy (%) | Execution Time (in milliseconds) |
|---|---|---|---|---|
| ID3 | 173 | 130 | 75.145 | 47.6 |
| C4.5 | 173 | 130 | 75.145 | 39.1 |

Table 2. Results of Singular Evaluation.

| Algorithm | Total Students | Students whose results are correctly predicted | Accuracy (%) |
|---|---|---|---|
| ID3 | 9 | 7 | 77.778 |
| C4.5 | 9 | 7 | 77.778 |

Thus, for a total of 182 students, the average percentage of accuracy achieved in Bulk and Singular Evaluations is approximately 75.275.

### ACKNOWLEDGEMENTS

We express sincere gratitude to our project guides, Ms M. Kiruthika, and Ms Smita Dange for their guidance and support.